**Title page**

Pedestrian collision avoidance in hemianopia during natural walking in immersive virtual reality

Jonathan K. Doyon[1], Sujin Kim[1], Alex D. Hwang[1], Jae-Hyun Jung[1,2]*

[1]Schepens Eye Research Institute of Massachusetts Eye and Ear, Department of Ophthalmology, Harvard Medical School, Boston, MA

[2]NVIDIA, Westford, MA

*jaehyunj@nvidia.com

Key words: locomotion, visual impairment, collision detection, collision avoidance, virtual reality, homonymous hemianopia

Word count: 8483

Funding information: NIH/NEI R01-031777, P30 EY003790

Disclosures: **J.K. Doyon**, None; **S. Kim**, None; **A.D. Hwang**, None; **J-H. Jung**, None

**Abstract**

Homonymous hemianopia (HH) patients report difficulties in avoiding collisions with other pedestrians. We evaluated pedestrian collision detection and avoidance behaviors in HH patients and healthy controls using a novel virtual reality (VR) walking with pedestrians, which enables natural walking behavior in an empty real-world corridor while viewing an immersive VR environment (shopping mall with colliding and other pedestrians) presented in a head-mounted display (HMD). Critically, it measures avoidance maneuvers in addition to collision detection.

Colliding and non-colliding pedestrian scenarios were developed for Meta Quest 2 using Unity. Ten normal vision (NV) subjects and 12 HH subjects detected and avoided collisions with virtual approaching and overtaken pedestrians initialized at bearing angles of ±20°, ±40°, and ±60°, with planned time-to-collision of 6 seconds in each trial.

HH subjects were less likely to detect and more likely to collide with pedestrians than NV, particularly for blind-side targets. Response times did not differ between groups but were faster for overtaken pedestrians. HH subjects also biased their head rotations toward the blind side and more after detection compared to before.

Collision avoidance difficulties as reported by HH subjects, which clinical measures fail to capture, were recorded and analyzed with objective measures. These metrics may offer further insights into the underlying mechanisms driving collision avoidance behaviors.

Our HMD-VR collision detection and avoidance paradigm enables natural walking behaviors and offers an affordable, objective assessment tool that may be adopted by clinicians for mobility enhancement and rehabilitation.

## 1. Introduction

Peripheral field loss (PFL) is a common consequence of various eye diseases, neurological disorders, and injuries[1,2]. Individuals with PFL face an increased risk of colliding with other pedestrians[3], greater difficulty in street crossings[4], and frequently encounter challenges in navigating and avoiding other obstacles while walking[5–9]. These difficulties significantly reduce mobility and can detrimentally affect the independence and quality of life of patients with PFL[6,10–14].

Current mobility assessments for low vision patients typically involve either self-reported measures[15,16] or more objective measures like preferred walking speed (PWS), collision rates, and orientation errors[17,18]. However, such assessments typically fail to characterize locomotor difficulties in avoiding dynamic obstacles (e.g., moving pedestrians or vehicles) in complex and cluttered environments (e.g., shopping malls, airports, or train stations). Indeed, outside of the lab, patients must discern the trajectories of and possible collisions with moving obstacles while detecting pathways that afford safe passage, a process which requires continuous environmental monitoring, accompanied by dynamic locomotor adjustments based on avoidance maneuvers, such as alterations in speed, path, and gaze direction[19]. Therefore, clinical mobility assessments must also evaluate collisions with dynamic obstacles in complex environments. However, creating such environments and managing dynamic collision scenarios as in the real-world presents significant challenges. Foremost, staging real collision events for PFL patients with sensory deficits carries potential risks of physical injury. Additionally, controlling real-world dynamic obstacles requires significant coordination efforts[20,21].

To avoid challenges related to physical injury risks and the control of obstacle dynamics, numerous mobility tests for patients with field loss have been introduced using VR technologies[22–26]. Given the virtual nature of VR obstacles, they pose no physical harm to the patients when interacting (e.g., when colliding or tripping). Obstacle scenarios (e.g., locations and movements) can also be controlled with precision timing and adjusted for randomization to avoid potential learning effects. However, many of these VR studies used virtual mazes with static obstacles only[27]. While these methods provide more controlled, safe, and variable environments than physical obstacle courses, they still do not include dynamic obstacle configurations. Critically, if the VR scenario lacks authenticity by using only static obstacles, the derived walking performance for patients with PFL may not generalize to real-world challenges[28].

Previously, we developed several realistic VR walking scenarios with various dynamic pedestrians to capture objective outcomes representing genuine walking performance in PFL patients[29–34]. These studies simulated walking on predefined paths through a busy virtual shopping mall and presented representations of colliding and non-colliding pedestrians utilizing looming cues (i.e., angular expansion) on a large television or wide field of view (FoV) display[29,31]. Subjects were required to detect and respond to the oncoming collision with a button press corresponding to the side of the visual field from which the pedestrian approached. To simulate the collision, a virtual pedestrian appeared at various locations and then walked toward the subject while other non-colliding pedestrians passed by. Our results demonstrated that patients with PFL have significantly lower detection rates on the blind side than normal vision (NV) subjects[29,31]. While these VR simulations successfully incorporated dynamic elements, they still differ from real-world walking in several aspects. First, participants in these simulations observed videos of simulated self-motion on a screen and detected oncoming collisions while sitting or standing. Importantly, the perceptual and cognitive load for detecting and avoiding oncoming hazards and controlling bodies during physical walking are more complex than standing and watching videos, and thus they might result in different behaviors. Second, while the changes in the looming effect (i.e., angular size) of the colliding pedestrian was simulated correctly, the fixed distance between the participant and the physical screen (i.e., lack of depth perception) could influence the perception of oncoming hazards since the colliding pedestrian cannot approach closer than the physical distance of the screen. Lastly, existing VR studies have predominantly focused on early hazard detection rather than collision avoidance, which is likely of greater consequence to the PFL patients. Therefore, those large screen walking simulations might affect the behaviors of PFL patients in detection, and thus new evaluation method that enables natural walking behavior for collision detection and avoidance as real-world is required.

In this paper, we present a more realistic evaluation of walking performance in PFL patients, specifically with homonymous hemianopia (HH), in which subjects physically walk and detect and avoid collisions with virtual pedestrians in a busy virtual shopping mall shown in a lightweight, standalone stereoscopic head-mounted display (HMD). Because the subject is able to walk and move head and eyes, the visual, attentional, and vestibular sensory dynamics more closely reflect those at work during real-world walking. Additionally, the stereoscopic HMD provides realistic depth cues, which results in a more immersive virtual world that allows other pedestrians to approach much closer to the subject in depth. Critically, these improvements allow the measurement of collision avoidance behaviors (e.g., slowing, turning, or passing) and head scanning in addition to detection rates and response times, resulting in a more complete assessment of naturalistic mobility performance for PFL patients.

Our primary hypotheses were that HH subjects would (1) detect fewer colliding pedestrians than NV subjects, (2) detect the colliding pedestrian later than NV subjects, and (3) make more collisions (i.e., contacts) with colliding pedestrians than NV subjects. Further, we expected these effects to be more pronounced for colliding pedestrians initialized at bearing angles on the blind side of the HH subjects. Additionally, we hypothesized that HH subjects may bias their horizontal head rotations (i.e., scanning behavior) toward their blind side while NV subjects would show no horizontal bias.

**2. Methods**

*2.1. Subjects*

Ten NV (5 women, $M_{age}$ = 34.10 years, SD = 6.86) and 12 HH (5 women, $M_{age}$ = 46.2 years, SD = 17.2) subjects were recruited from the Schepens Eye Research Institute of Massachusetts Eye and Ear (Boston, MA). The HH group was significantly older (Welch's t(14.9) = -2.23, p = 0.04, d = -0.92) but did not differ in gender distributions (Fisher's exact test, p = 1.00). The two groups differed in terms of self-identified race, where the NV group included 6 Asian and 4 White subjects, compared to the HH group which included 1 Black subject, 1 Asian subject, and 10 White subjects (Fisher's exact test, p = 0.02); all subjects identified as non-Hispanic ethnicities. Subjects provided written informed consent before participating. The study was approved by the Institutional Review Board (protocol: 2021P003330) as part of a clinical trial (ClinicalTrials.gov identifier: NCT05141604) and conducted in accordance with the Declaration of Helsinki.

*2.2. Scenario development*

All scenarios and aspects of the testing environment were programmed using the Unity game engine (Unity Technologies, San Francisco, CA), which was then ran on a Meta Quest 2 (Meta, Menlo Park, CA) HMD. Full development technical details and scenario design can be found in our previous work[35]; details relevant to the current study are summarized below.

We modeled three different types of pedestrians[33,34]: approaching pedestrians (Fig. 1a), overtaken pedestrians (Fig. 1b), and non-colliding pedestrians (i.e., distractors). The bearing angle (i.e., the angle between the paths of the subject and the colliding pedestrian; β) is considered relative to the subject's heading, such that the bearing angle that coincides with the subject's heading direction is $β = 0°$, rightward angles are positive, and leftward angles are negative. An assumed collision point is placed at a predetermined location along the subject's path at a distance dependent on the subject's PWS and a designed time-to-collision (TTC). The colliding pedestrian is then set to walk toward the collision point.

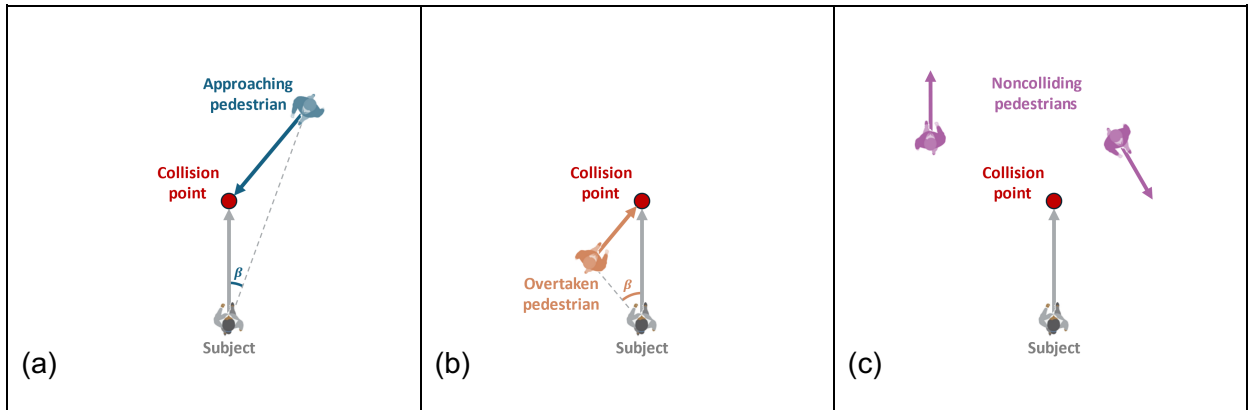

**Fig. 1. Modeling of different types of pedestrians.** (a) Approaching (face-to-face) pedestrian with bearing angle β = +20°, (b) overtaken (side-to-side) pedestrian with bearing angle β = -40°, (c) multiple non-colliding pedestrians at various bearing angles.

In the approaching case (i.e., face-to-face collision course, Fig. 1a), the pedestrian is placed at a location beyond the collision point at a bearing angle β with respect to the subject[30,32]. We set the same distances to the collision point from the subject and the approaching pedestrians, and the walking speed of the approaching pedestrian was the same as PWS of the subjects. In the overtaken case (i.e., side-to-side collision course, Fig. 1b), the pedestrian is placed at a location between the subject and the collision point (and consequently, closer and more visible to the subject), also at a bearing angle β with respect to the subject. To keep the larger than 45°, the distance to the collision point from the overtaken pedestrians was closer than the subjects. Note that in order for the subject to collide with the overtaken pedestrian, the pedestrian must be walking at a slower speed than the subject.

Importantly, in the real world, the two types of colliding pedestrians differ in two key dimensions—first, in the division of the social responsibility for avoiding collisions. For two normally sighted approaching pedestrians, since both the pedestrians can clearly see each other, the responsibility to avoid is shared between both parties. For the normally sighted pedestrian that is overtaken by the other, the responsibility is carried mostly by the overtaking pedestrian since they fall outside of the overtaken pedestrian's visual field. However, when PFL patients are unknowingly overtaking the other pedestrian (who's outside their limited visual field), the collision is more likely to happen due to both the overtaking and overtaken pedestrians being located in each other's' blind fields and thus failing to detect the collision. The other key difference is in the relative walking speeds of the two pedestrians. Due to the opposing alignments of heading directions in the approaching scenario, both of the pedestrians' walking speeds are faster relative to one another. Oppositely, because the heading directions are in closer alignment in the overtaken scenario, the relative walking speeds of the pedestrians are slower.

*2.3. Procedure*

Two experimenters were present during all stages of the data collection process. Each experimental session began with a short verbal overview of the requirements and structure of the task. The experimenters then used the HMD's casting feature to wirelessly stream the HMD's display to a nearby laptop. One experimenter then wore the HMD and completed a set of example trials while the subject observed, asking questions if needed. The subject's interpupillary distance was measured so that the HMD could be adjusted for clarity. The eye height was also measured to match the viewpoint in the VR space with subject's standing viewpoint in the real-world. The subject then completed an acclimation scenario and a PWS measurement scenario before beginning the main scenario with the collision detection and avoidance task.

*2.3.1. Acclimation scenario and alignment task.* The acclimation scenario consisted of a series of walking trials in which the subject physically walked in the virtual mall without any pedestrians present. The purpose of this scenario was first to help the subject acclimate to the virtual environment, and second to introduce the alignment task, which was used to set the walking path on each trial. During the alignment task, the HMD's video passthrough function was activated, which captured the physical space outside of the HMD and superimposed it onto the virtual world such that both were simultaneously visible (Video S1). The subject then aligned the virtual world space with the physical world space by aligning the intended walking path in the virtual world with a physical strip of tape affixed to the floor of the testing space. This procedure ensures that the subject maintains a safe pathing orientation that coincides with the longer dimension of the physical room for the potential path changes avoiding collisions.

*2.3.2. PWS measurement scenario.* The PWS measurement scenario consisted of a series of trials in which the subject walked until reaching an invisible trigger-point (approximately 3 meters) which made visible only the non-colliding pedestrians (i.e., no collisions would occur in this scenario). The subject was instructed to walk as naturally as possible and encouraged to look around as they pleased with the understanding that no collisions would occur. The trial ended once the subject reached a second invisible trigger point (approximately 10 meters) signaling the end of the trial. We measured their walking speed on each trial and computed the average across 12 total trials to derive each subject's PWS. This value was then used to determine the location of the assumed collision point in the main scenario based on a designed TTC of 6 seconds.

*2.3.3. Main walking scenario with collision detection and avoidance task.* The main scenario consisted of 32 randomized walking trials: 20 "collision" trials which involved a colliding pedestrian among 10 other non-colliding pedestrians, and 12 null trials in which no colliding pedestrian was present (i.e., only the non-colliding pedestrians were present). On the collision trials, a single approaching (Fig. 2a) or overtaken (Fig. 2b) colliding pedestrian walked toward the collision point from a location at one of 6 initial bearing angles: ±20°, ±40°, and ±60°; note that approaching pedestrians could only be initialized at ±20° or ±40° due to the restrictions detailed above. Because the HMD's FoV is smaller than ±60° (horizontally ~90°), pedestrians initialized at ±20° are visible without head scanning in NV subjects, while HH may detect with sufficient gaze scans into the blind-side); pedestrians initialized at ±40° are near the HMD FoV limits and may be detected in NV without head scanning, but blind-side detection without gaze movement is not possible for HH; and pedestrians initialized at ±60° cannot be detected without head rotation for both NV and HH subjects. Each pedestrian walked in a straight path with constant speed;

assuming the subject walks at a similar constant speed (or slightly faster than the pedestrian when overtaking), the two will collide at the collision point after the designed TTC has elapsed.

Once the subject detected a colliding pedestrian, they responded by pressing a button on the Quest 2 controller that corresponded to the side of the colliding pedestrian (i.e., left or right from the path). After responding, the subject then naturally avoided the colliding pedestrian as they would in any other real-world situation (e.g., speed and/or path change). Collision events were recorded as any overlapping instances of the areas spanned by the subject's and pedestrian's shoulders (both fixed at 25 cm radius from center).

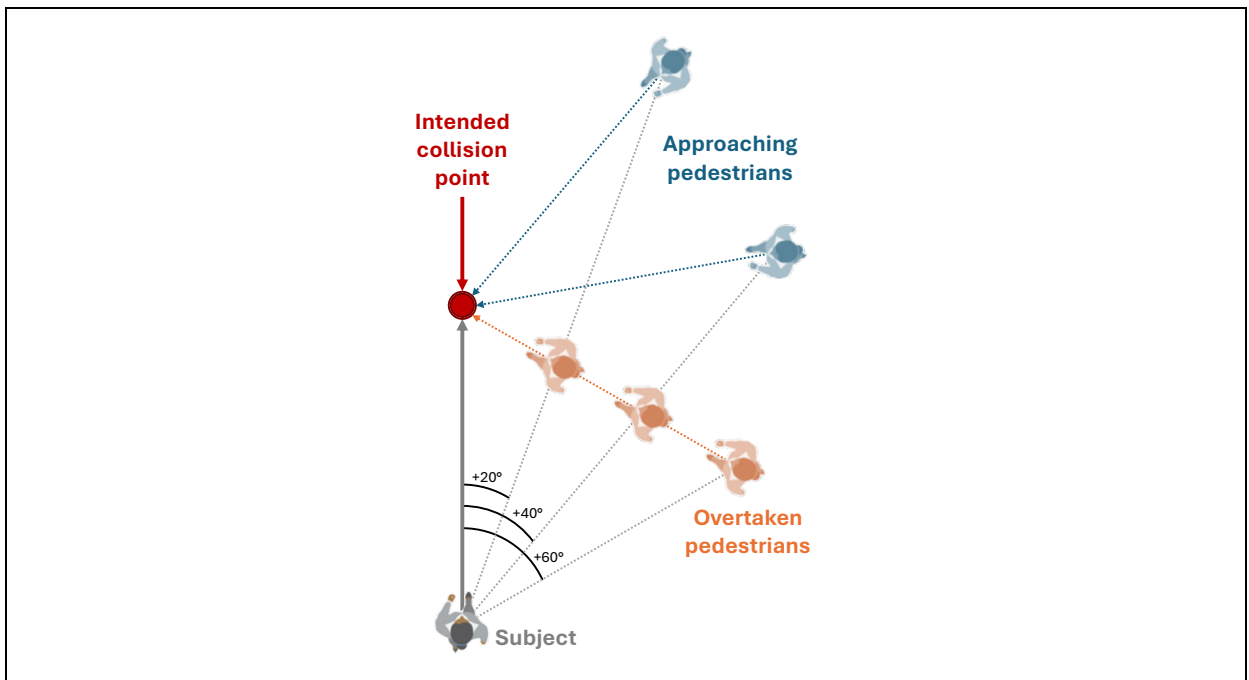

Fig. 2. Schematic of two types of colliding pedestrians based on a straight path: Approaching (i.e., face-to-face; blue) pedestrian collision from 20° and 40° and Overtaken pedestrian collision (i.e., side-to-side; orange) from 20°, 40°, and 60°. In each collision event, a colliding pedestrian appears at the pre-defined position (e.g., 2m away from the subject for +20° condition) and walks toward the intended collision point. Note that pedestrian positions were symmetrical across the visual field, so only positive bearing angles (i.e., the right half of the visual field) are depicted for simplicity.

*2.4. Measurements*

All measurements and behaviors were logged in real time by the HMD and handheld controllers. We recorded head position (x, y, z coordinates) and rotation (φ, θ, ψ Euler angles), walking speed (m/s), collision detection rate, response time (s), collision rate, and FoV (Fig. 3; Video S2). Response times were considered the elapsed time between the appearance of the pedestrians and the button press response.

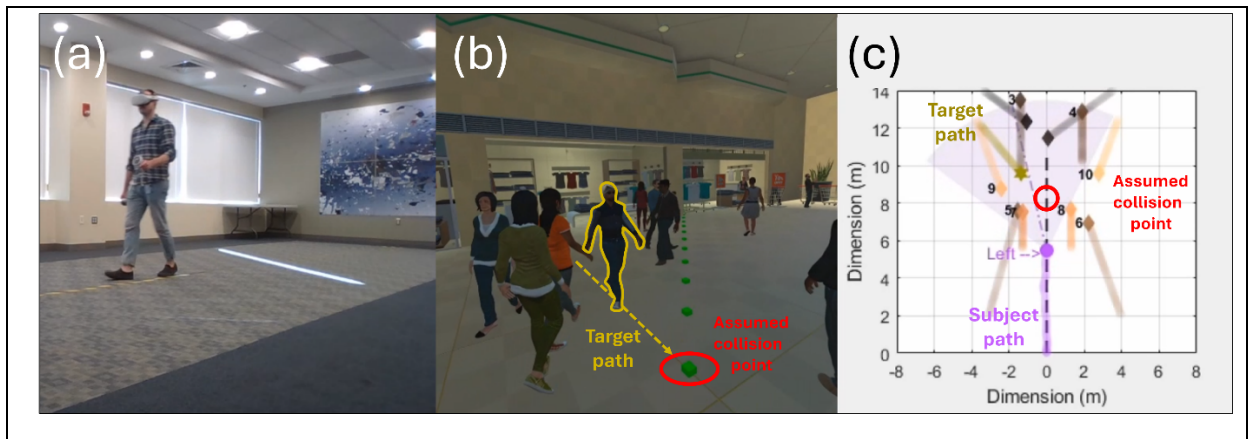

Fig. 3. HMD VR collision detection and avoidance task (s*ee the attached Video S2*). (a) A subject walks on a real corridor while wearing the HMD. (b) The HMD shows a predefined walking path (green floor markers) in a busy VR shopping mall with multiple pedestrians. (c) Walking trajectories of the subject (purple), colliding pedestrian (green), and non-colliding pedestrians (various shades of orange and gray), and horizonal FoV (purple wedge).

*2.5. Analysis*

We modeled detection rates, response times, collision rates, and head rotations with a series of linear mixed effects regression models[36]. In each case, we used forward stepwise selection to determine the most parsimonious model for our study design. The dependent variable was modeled as a non-linear combination of the fixed effects of group (NV vs. HH), pedestrian type (approaching vs. overtaken pedestrians), and initial target bearing angle (±20°, ±40°, ±60°). To account for individual differences, we specified random intercepts for each subject.

For each analysis, we begin by fitting the model containing only the effect of group and the random intercept. We then fit a second model which adds the effect of pedestrian type, then test for a significant improvement in the loglikelihood of the model, which represents our criterion for whether the added effect should be retained or removed. We continue this procedure until all effects have been entered into the model. Finally, due to the asymmetric nature of the field loss in HH (i.e., only one hemifield is visible), we compared performance for HH subjects as a function of their blind vs. seeing sides by binarizing the target pedestrian's initial bearing angle; this is done by coding the initial target bearing angle factor as having started from an initial bearing angle less than zero (blind side) or greater than zero (seeing side). Note that we standardized target initial bearing angles across left and right HH subjects prior to any analysis such that bearing angles and rotations less than zero correspond to targets in, and scans toward the blind-side, respectively.

## 3. Results

*3.1. PWS*

Average PWS of HH subjects was slightly slower (M = 0.9 m/s, SD = 0.09) than NV subjects (M = 0.98 m/s, SD = 0.12), but the difference was not significant (std. β = -0.77, p = 0.07, 95% CI = [-1.61, 0.08]) (Fig. 4).

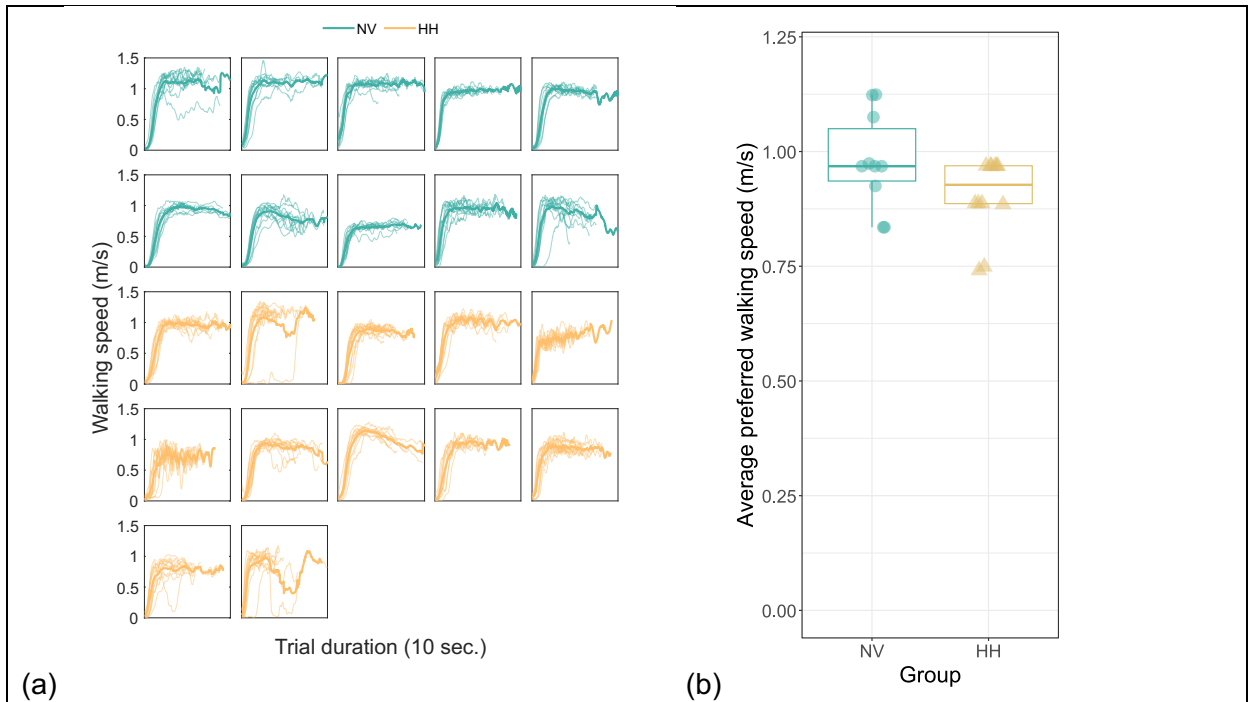

Fig. 4. Average PWS: individual trials plotted as curves (mean represented by the darker curve) for (a) each subject and (b) group level comparison. The upper and lower hinges of the boxplots correspond to the first and third quartiles and the upper and lower whiskers extend to 1.5 times the inter-quartile range. The notches of the boxplots correspond to a 95% confidence interval around the median.

*3.2. Collision detection*

Though the collision detection failures (misses) were rare, HH subjects made significantly more misses (50 out of 439) than NV (8 out of 439; $\chi^2(1)$ = 23.58, p < .001) (Fig. 5a). To evaluate the percentage of collisions across subject groups, pedestrian types, and initial target bearing angles, we fit a logistic mixed effects regression model which included random intercepts for each subject to account for individual differences. Pedestrian type did not significantly improve model fit beyond the group model (ΔAIC = -0.53, $\chi^2(2)$ = 3.47, p = 0.18) and was removed. Initial target bearing angle significantly improved model fit (ΔAIC = 6.11, $\chi^2(2)$ = 10.10, p = 0.01) and was retained, resulting in the final model (conditional $R^2$ = 0.42, marginal $R^2$ = 0.30).

The analysis showed a significant negative main effect of group (std. β = -2.16, p = 0.002, 95% CI = [-3.49, -0.83]) indicating that the HH subjects were less likely to detect the colliding pedestrian than NV subjects (Fig. 5a). The negative main effect of initial target bearing angle was also significant (std. β = -1.07, p = 0.03, 95% CI = [-2.03, -0.10]), indicating that the likelihood of detecting the colliding pedestrian was lower for targets that were initialized at larger bearing angles on the blind side than the seeing side of the visual field in HH subjects (c.f., left side and right side in NV subjects, respectively). These effects were superseded by a significant positive interaction between group and initial target bearing angle (std. β = 1.37, p = 0.01, 95% CI = [0.36, 2.38]), which indicated a crossover effect in which HH subjects were less likely to detect targets initialized at larger bearing angles in the blind side for HH while NV subjects were (slightly) less likely to detect targets initialized at farther peripheries in the right side.

For HH subjects, there were no significant effects specific to the blind side (blind vs. seeing side: std. β = -0.41, p = 0.92, 95% CI = [-0.76, -0.05]; initial target bearing angle (absolute value): std. β = -0.18. p = 0.17, 95% CI = [-0.43, 0.08]; interaction: -0.15, p = 0.42, 95% CI = [-0.51, 0.21]), indicating that HH subjects were equally likely to detect the colliding pedestrian whether they were located in the blind side or seeing side, regardless of the initial target bearing angle (Fig. 5b).

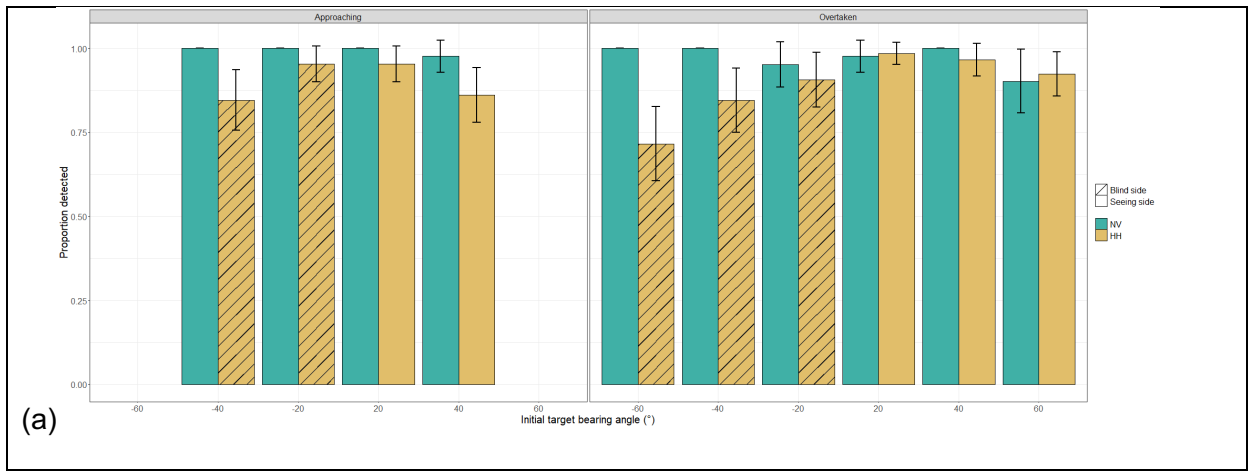

(a)

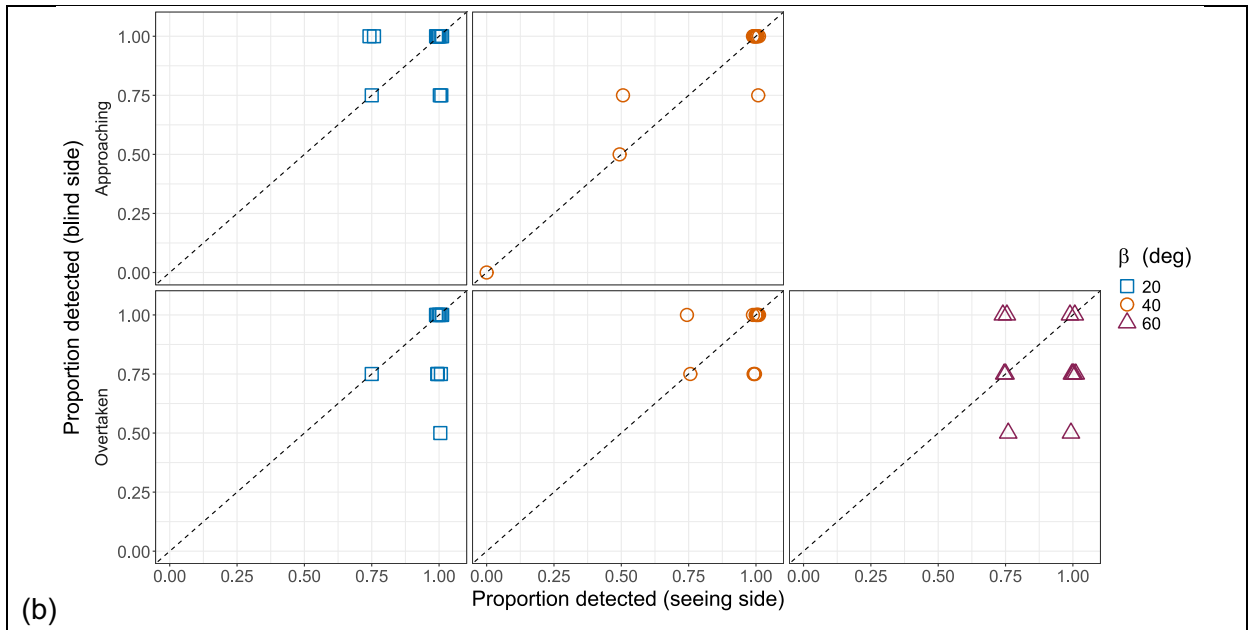

(b)

Figure 5. Collision detection rate (a) as a function of subject group and initial target bearing angle, error bars correspond to 95% confidence intervals, (b) subject-level data (HH only) depicting proportion of targets detected on the blind side vs. the seeing side as a function of initial target bearing angle (β); data are jittered slightly to show overlaid points (b).

## 3.3. Response times

To evaluate response times to detecting the colliding pedestrian across subject groups, pedestrian types, and initial target bearing angles, we fit a linear mixed effects regression model which included random intercepts for each subject. We first removed incorrect responses (e.g., right-side response to a left-side target), false alarms, and responses that occurred after a collision occurred. We then log-transformed the response time measurements to normalize its distribution.

Group did not significantly improve model fit beyond the intercept (ΔAIC = -0.6, $\chi^2(1)$ = 1.39, p = 0.24) and was removed from the model. Pedestrian type significantly improved model fit (ΔAIC = 235.69, $\chi^2(1)$ = 237.69, p < 0.001) and was retained. Initial target bearing angle did not significantly improved model fit (ΔAIC = 1.68, $\chi^2(2)$ = 5.68, p = 0.06) and was removed, resulting in the final model (conditional $R^2$ = 0.46, marginal $R^2$ = 0.19). The main effect of pedestrian type was significant (std. β = -0.89, p < 0.001, 95% CI = [-1.00, -0.78]) indicating that response times were faster for overtaken pedestrians than for approaching pedestrians (Fig. 6a).

For HH subjects, there were significant main effects of both blind vs. seeing side (std. β = 0.26, p = .045, 95% CI = [-0.05, 0.57]) and of the initial target bearing angle (std. β = -0.36, p = .002, 95% CI = [-0.58, -0.14], indicating that response times were slower for pedestrians on the blind side and for pedestrians initialized at larger bearing angles. These effects were superseded by a significant interaction between the two factors (std. β = 0.45, p =.005, 95% CI = [0.14, 0.76]), indicating a crossover effect such that response times were faster for targets in the blind side at smaller bearing angles, but slower for farther periphery (larger bearing angle) targets (Fig. 6b).

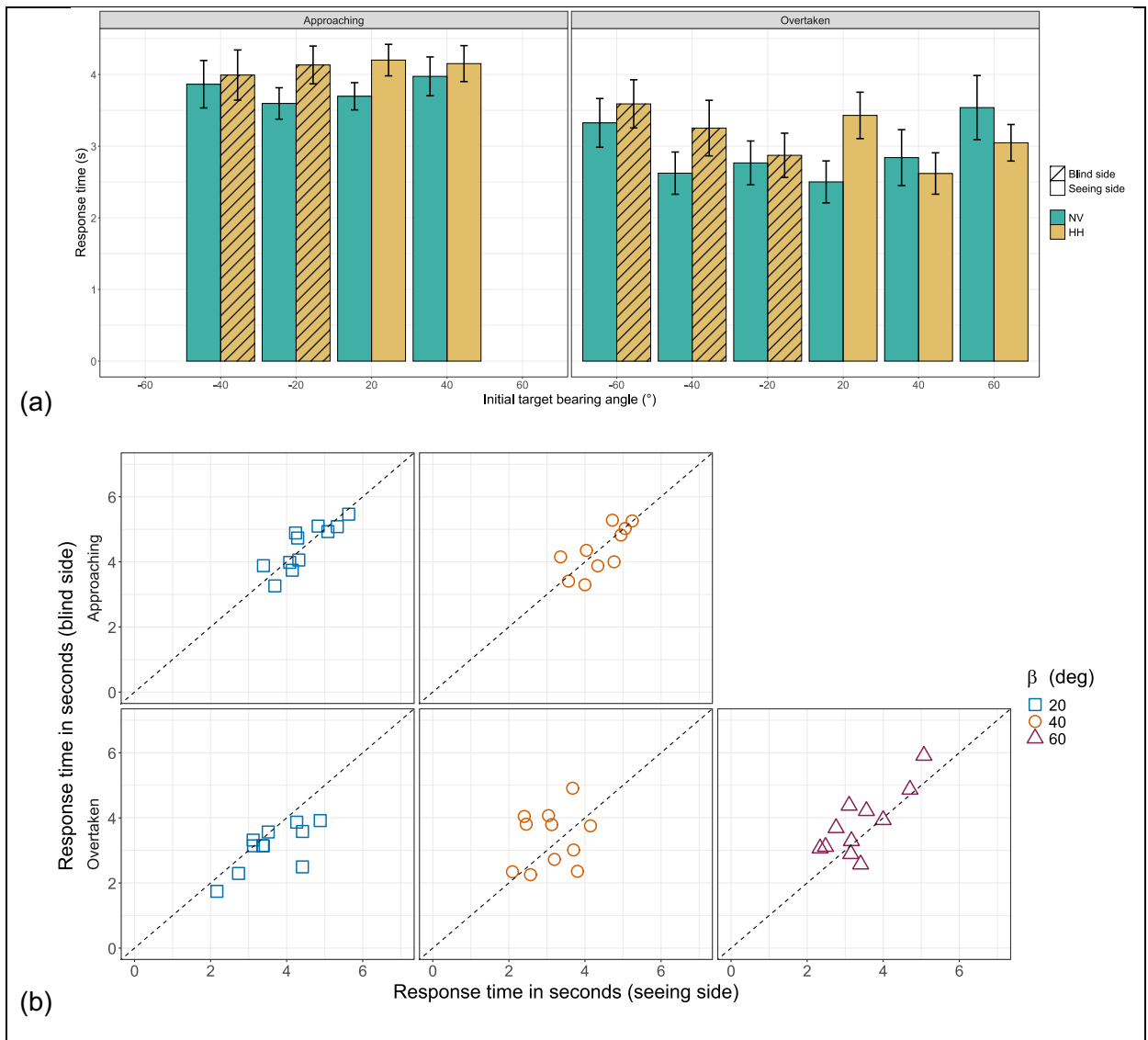

Figure 6. Response times of the collision detections (a) as a function of subject group and initial target bearing angle, error bars correspond to 95% confidence intervals, (b) Subject-level data (HH only) comparing collision detection response times to pedestrians initialized in the blind side vs. the seeing side as a function of initial bearing angle.

*3.4 Collision rates*

In general, collisions with pedestrians were rare (only 38 out of 878 total trials; 4.3%). Of these collisions, HH subjects made significantly more collisions (30 out of 439) than NV (8 out of 439; $\chi^2(1) = 8.45$, p = 0.004). To evaluate the likelihood of the subject colliding with the pedestrian across subject groups, pedestrian types, and initial target bearing angles, we fit a logistic mixed effects regression model with subject as a random effect.

Pedestrian type did not significantly improve model fit beyond the effect of group (ΔAIC = -3.82, $\chi^2(2)$ = 0.17, p = 0.92) and was removed from the model. Initial target bearing angle significantly improved model fit (ΔAIC = 12.69, $\chi^2(2)$ = 16.681, p < 0.001) and was retained, resulting in the final model (conditional $R^2$ = 0.26, marginal $R^2$ = 0.19).

The analysis showed a significant main effect of group (std. β = 1.15, p = 0.04, 95% CI = [0.06, 2.23]) indicating that HH subjects were more likely to collide with a pedestrian (Fig. 7a). The main effect of initial target bearing angle was not significant (std. β = 0.66, p = 0.11, 95% CI = [-0.15, 1.46]), but its interaction with group was significant (β = -1.42, p = 0.002, 95% CI = [-2.34, -0.51]) indicating that HH subjects were more likely to collide with the pedestrian when initially appeared in the blind side (i.e., when β < 0), while NV subjects were more likely to collide with the pedestrian when initialized in the right side (i.e., when β > 0).

For the HH subjects, there was a significant interaction between the blind vs. seeing sides and the initial target bearing angle (std. β = 0.19, p = 0.03, 95% CI = [0.02, 0.36]), indicating that HH subjects made more collisions with pedestrians on the blind side, particularly when those pedestrians were initialized at farther periphery (larger bearing angles) (Fig. 7b). Neither of the constituent main effects were significant (blind/seeing side: std. β = 0.28, p = 0.43, 95% CI = [0.10, 0.45]; initial target bearing angle: std. β = 0.08, p = 0.21, 95% CI = [-0.04, 0.20]).

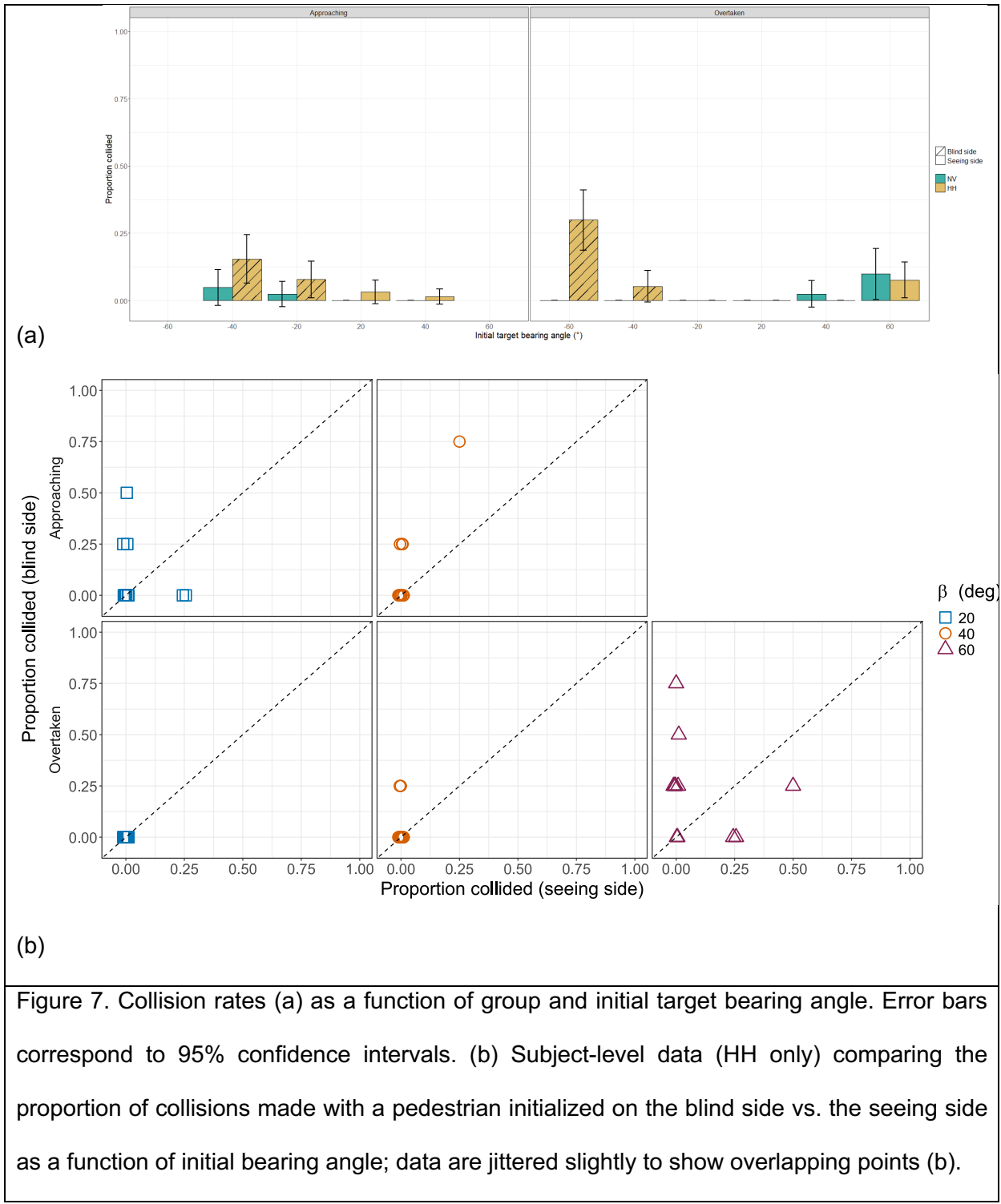

Figure 7. Collision rates (a) as a function of group and initial target bearing angle. Error bars correspond to 95% confidence intervals. (b) Subject-level data (HH only) comparing the proportion of collisions made with a pedestrian initialized on the blind side vs. the seeing side as a function of initial bearing angle; data are jittered slightly to show overlapping points (b).

*3.5 Head rotations.*

To evaluate whether HH subjects show any bias in their head rotations (i.e., scanning behavior), particularly before and after detecting the potential collision, we computed the average horizontal and vertical head rotations of the HMD relative to the walking path as an index of scanning bias toward the left side (blind side in HH) vs. the right side, or as an index of looking upward vs. downward. Means were compared against zero (i.e., no bias) using one-sample t-tests with Holm-Bonferroni corrected p-values. We then fit a mixed effect linear regression with subject as a random factor to investigate the relationships between group and before/after detection.

*3.5.1 Horizontal rotations.*

HH subjects showed a significant bias in head scanning towards their blind side both before ($t(11) = -3.29$, $p = 0.02$, Cohen's $d = -0.96$, 95% CI = [-1.63, -0.25]) and after the detection ($t(11) = -3.73$, $p = 0.01$, Cohen's $d = -1.04$, 95% CI = [-1.73, -0.31]) (Fig. 8). NV subjects showed no bias in their head scanning toward either side both before ($t(9) = -1.41$, $p = 0.18$, Cohen's $d = -0.55$, 95% CI = [-1.21, 0.13]) and after ($t(9) = -2.01$, $p = 0.15$, Cohen's $d = -.59$, 95% CI = [-1.25, 0.10]. The regression analysis (conditional $R^2 = 0.62$, marginal $R^2 = 0.20$) showed no significant effect of group (std. $\beta = -0.36$, $p = 0.36$, 95% CI = [-1.16, 0.43] or before/after detection (std. $\beta = 0.16$, $p = 0.58$, 95% CI = [-0.42, 0.73]). However, their interaction was significant (std. $\beta = -0.82$, $p = 0.04$, 95% CI = [-1.60. -0.05]) indicating that the effect of before/after detection was stronger in the HH group than in the NV group.

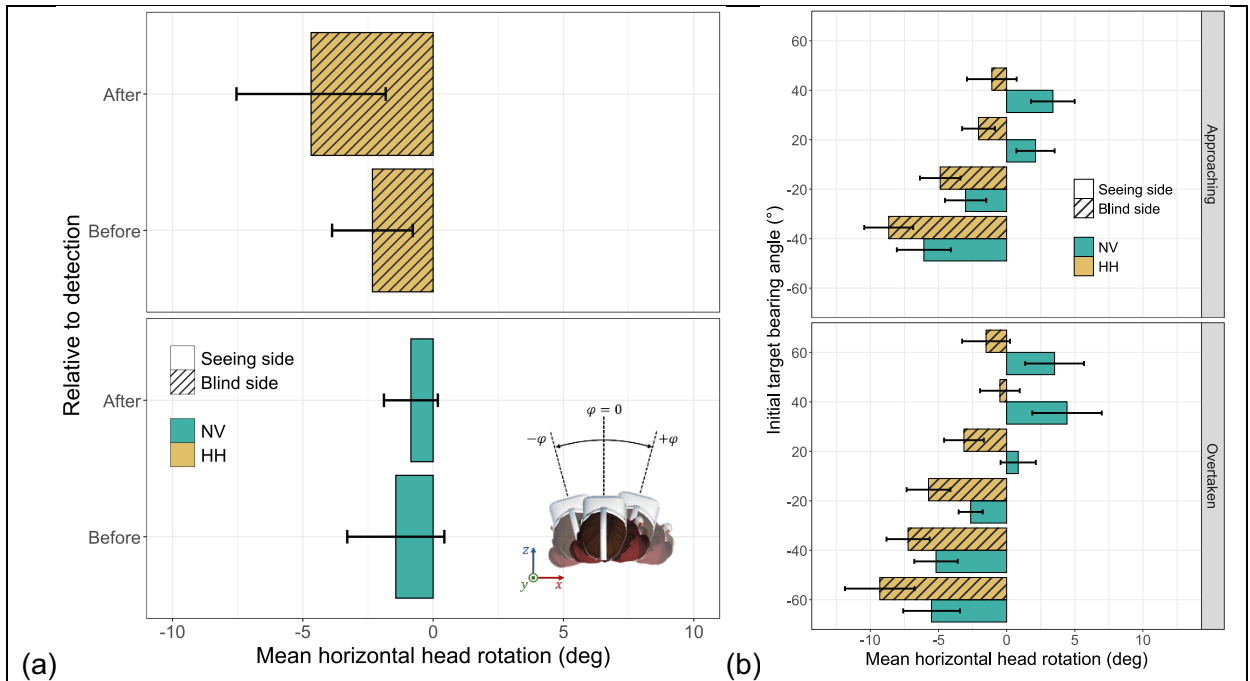

Fig. 8. Mean horizontal head rotation angles (φ) (a) before detection (i.e., during search) and after detection (i.e., during avoidance; a). (b) Mean horizontal head rotations as a function of initial target bearing angle and pedestrian type. Hatched bars (negative sign) indicate blind-side for the HH field loss group. Error bars represent 95% confidence intervals.

*3.5.2 Vertical rotations.*

HH subjects showed a significant upward bias in vertical head rotation both before ($t(11) = 4.68$, $p = .002$, Cohen's $d = 1.35$, 95% CI = [0.54, 2.13]) and after ($t(11) = 6.46$, $p < .001$, Cohen's $d = 1.86$, 95% CI = [0.89, 2.80] the detection (Fig. 9), while NV subjects showed a similar bias only after the detection ($t(9) = 3.49$, $p = .01$, Cohen's $d = 1.10$, 95% CI = [0.29, 1.88], not before ($t(9) = 1.31$, $p = 0.22$, Cohen's $d = 0.42$, 95% CI = [-0.24, 1.05]. The regression analysis (conditional $R^2 = 0.95$, marginal $R^2 = 0.19$) showed no significant effect of group (std. $\beta = 0.68$, $p = .10$, 95% CI = [-0.13, 1.48], indicating HH and NV subjects showed similar vertical head rotations. There was a significant effect of before vs. after detection (std. $\beta = 0.74$, $p < .001$, 95% CI = [0.54, 0.95]) indicating that there was a significant upward bias in head rotation after the detection in both subject groups. The interaction between group and before vs. after detection was not significant (std. β = -0.18, p = .02, 95% CI = [-0.45, 0.10]).

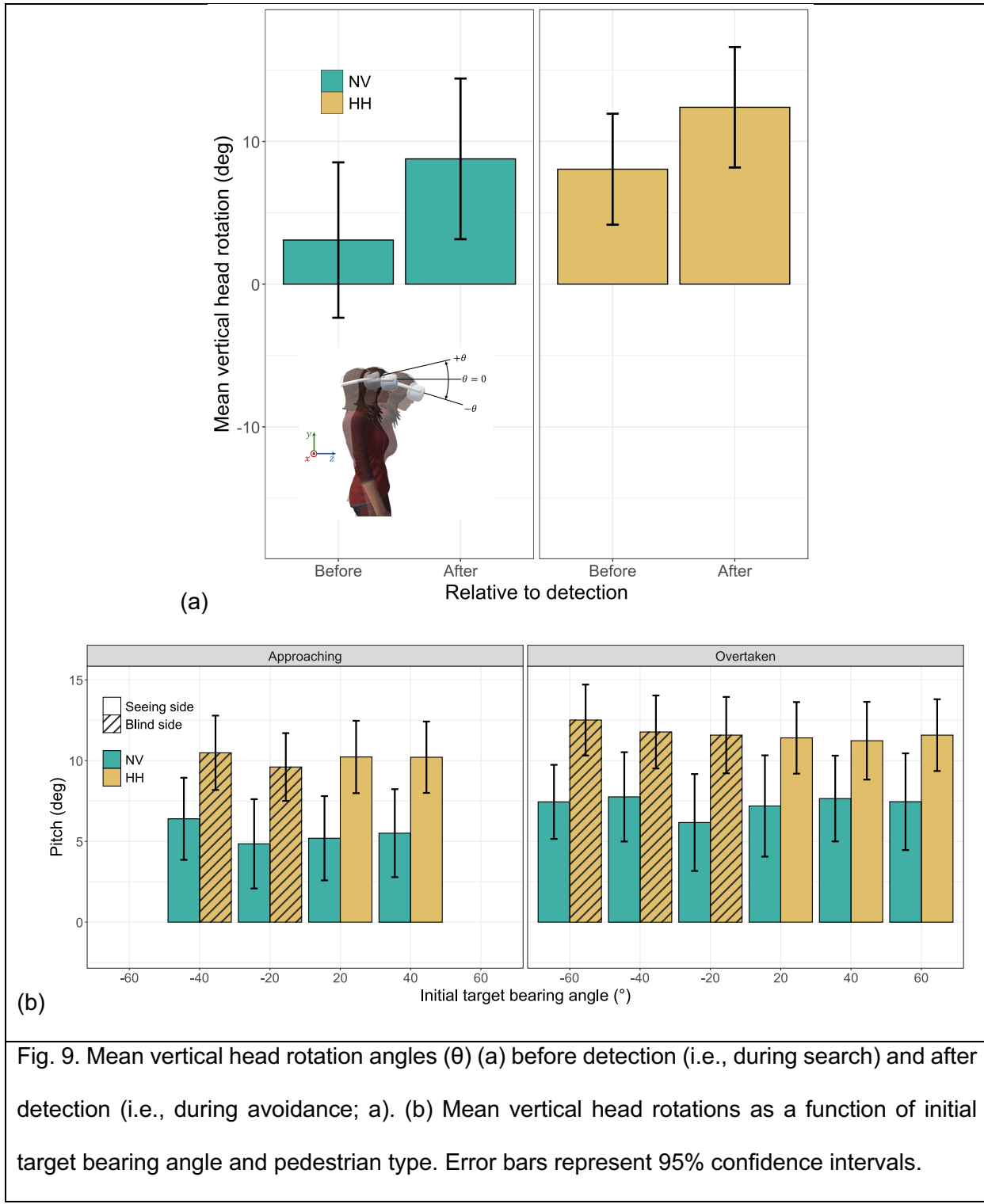

Fig. 9. Mean vertical head rotation angles (θ) (a) before detection (i.e., during search) and after detection (i.e., during avoidance; a). (b) Mean vertical head rotations as a function of initial target bearing angle and pedestrian type. Error bars represent 95% confidence intervals.

## 4. Discussion

We developed a novel platform that allows the assessment of several objective metrics of mobility performance, improving on common mobility assessments which typically rely on either self-report[15,16] or highly restricted testing scenarios[22–26,28]. Critically, our testing platform allows the subject to walk freely with unrestricted head and eye scans, which provides more realistic scenarios in which to evaluate mobility difficulties (e.g., collision avoidance) faced by PFL patients in everyday life[3–9] enabling their natural walking behaviors. We evaluated collision avoidance performance and head rotation behaviors in groups of NV and HH subjects to determine whether HH subjects would show deficits in detecting and avoiding virtual colliding pedestrians across a number of approaching and overtaken configurations.

*4.1. Homonymous hemianopia subjects detect most collisions, but fewer than normal vision.*

We found some support for our initial hypothesis that HH subjects would detect fewer colliding pedestrians than NV subjects. However, this effect must be qualified by the fact that both groups performed near or at ceiling (Fig. 5). HH subjects showed the lowest detection performance for targets initialized on their blindside, particularly for pedestrians requiring the largest magnitude of head scans (i.e., -60° pedestrians). Note that although 60° pedestrians are out of the HMD FoV and thus in the blind field of both NV and HH subjects, any small amount of head movement while walking easily bring the pedestrians into the visual field of NV. However, very excessive (~60°) and intentional head scanning is required in the case of HH subjects.

Even in the given condition and difficulties, the apparent ceiling effect may suggest that HH subjects do not have significant difficulty in detecting colliding pedestrians, particularly from the blindside, as has been reported in other investigations[29–32]—this is likely due to the unrestricted ability to scan into the blind side and enabled natural walking behavior in our setup, resulting in a higher likelihood of detection. Lower detection rates of farther periphery pedestrians in even seeing side of HH may indicate the compensatory gaze movement to cover the blind side and thus scarification of seeing side detection.

Additionally, the ceiling effect may simply result from the artificiality of the task design (see section 4.6). However, considering the consequences of real-world collisions with actual pedestrians (e.g., physical injuries, falls, altercations, etc.), even in only few, the observed detection rate deficits on the order of 10-20% may still represent significant risks to HH patients in real-world scenarios.

*4.2. Homonymous hemianopia subjects detect collisions as fast as normal vision subjects.*

We found no support for our initial hypothesis that HH subjects would be slower in detecting the colliding pedestrians compared to the NV subjects. We suspect this is due to the fact that compensatory scanning allows the HH subjects to successfully detect the colliding pedestrian by the same temporal margins as NV subjects, who presumably have no need for compensatory scanning (see Fig. 6 and section 4.4). It is also possible that the relatively straightforward navigation task (i.e., straight pass) contributed to the results. If subjects had more dynamic and realistic navigation tasks, such as finding a path, a goal, or checking terrain, which require more attention to the center and limit free scanning in time and magnitude, the results might have been different. Alternatively, this result may reflect another task constraint, that all collision events were designed with a 6 second TTC, which may permit more than enough time to search and detect the colliding pedestrian. Indeed, ongoing preliminary analysis of the timing of the observed collision avoidance behaviors[37] suggests that both HH and NV subjects initiate avoidance behaviors at or around 0.5-1 second TTC. If subjects then wait until just a second or so before the collision before initiating an avoidance maneuver, then the designed 6 seconds TTC may not be appropriate to capture the true detection time. However, a recent computational analysis[38] of pedestrian collision risk for visual field loss patients found that there may be little if any risk of collision for designed TTCs of 2 seconds or less. Ongoing work will empirically test such cases.

Our analysis did show an effect of pedestrian type (Fig. 6) such that response times were faster for overtaken pedestrians (i.e., shoulder-to-shoulder) than for approaching pedestrians (i.e., face-to-face). This is likely due to the differences in the pedestrian's radial distance and angular size. Note that HMD provided stereoscopic depth cues, ensuring that both distance and size cues of the pedestrians were accurately represented to be perceived. Specifically, overtaken pedestrians were initialized at a much closer radial distance between the subject and the assumed collision point and, consequently, subtending a larger visual angle than the approaching pedestrians, which were initialized beyond the assumed collision point with a much smaller visual angle. One possibility is that subjects may treat the detection task as a simple visual search task, sequentially sampling the various pedestrians accumulating evidence for the collision detection decision at varying rates such that the overtaken pedestrians provide stronger evidence at earlier samples than the approaching counterparts. Another possibility may be that the array of possibilities for safe passage (i.e., open spaces) is narrowed much faster in the overtaken case due to pedestrians larger size and closer positioning compared to the approaching case, in which the possibilities for safe passage narrow much later, or closer to the collision, allowing for later responding. Future work on such informational arrays (i.e., affordances for safe passing) as landscapes of attractors[39] is needed.

*4.3. Homonymous hemianopia subjects more collide with pedestrians than normal vision subjects, particularly with pedestrians initialized on the blind side.*

Similar to our analysis of the collision detection data, both subjects groups successfully avoided most collisions (~96% overall). This is also the case in real life, since we rarely encounter collision events while walking. However, HH subjects made significantly more collisions than NV subjects, especially with those colliding pedestrians initialized on the blind side—a finding that partially supports our initial hypothesis. However, the dearth of collisions in both groups, paired also with the apparent ceiling effect in the detection rates, suggests a potential floor effect in the observed collision rates that may be driven by the artificiality of the task. Interestingly, the NV group collided only with pedestrians initialized on the right side, potentially due to the cultural practice of walking (and driving) on the right side of path as is done in the United States. For example, when walking through a constrained path (e.g., sidewalk), if one walks along the right side of the path, then most collision risks will arise from the opposing flow of foot traffic which is present on the left side of the path, relative to the walker. This might result in an attentional bias toward the left side for right-side-walkers and an opposite bias in those who are socialized in regions that maintain left-side walking and driving.

Despite their apparent adeptness in avoiding collisions in the current task, HH subjects collided with nearly 20% of pedestrians at the farthest periphery (largest bearing angle) in the blind side (which happen to be overtaken pedestrians; see section 2.2). Given the potentially serious consequences of colliding with other pedestrians in crowded environments—particularly ones that cannot see the HH patient as in the overtaken case—1 in 5 collisions resulting in contact and a potentially serious fall or other physical injury may represent a serious risk and limit to mobility and independence. While the task used in the current study has many constraints that may not accurately reflect common real-world scenarios (see section 4.6), Kurukuti et al.[38] argue that pedestrians that are overtaken at 60° in the blind side present the greatest risk and are indeed the most realistic situation in the current task, as the subject is in the blind field of the pedestrian and vice versa. Whether the observed collision rate in the current task corresponds to the true rate of collisions for HH patients in the real world is an open question and further investigation.

*4.4. Homonymous hemianopia subjects bias their horizontal head rotation toward their blind side both before and after detecting the colliding pedestrian, while normal vision group does not.*

Our initial hypothesis found partial support such that HH subjects biased their horizontal head rotations toward their blind side (i.e., $\varphi < 0°$). However, the magnitude of this bias was quite small—roughly 3-5° (Fig. 6). This may be surprising considering the magnitude of head rotation necessary to place the least eccentric blind side colliding pedestrian (i.e., -20°). However, in a recent study[40] of HH subjects walking in the real world through busy sidewalks and street crossings, similar biases (1.9-3.3°) in HH subjects' gaze scans toward the blind side were observed. The authors argue that such compensatory scanning may reflect greater reliance on top-down guidance of visual attention (e.g., when actively searching for potential colliding pedestrians) as opposed to more bottom-up guidance (e.g., when selecting moment-to-moment passages or gaps that afford safe passing). While this may be a plausible explanation for the observed effect in our results, we cannot be certain without concurrent eye tracking data.

Additionally, the nature of the task presented here can be broken down into two distinct phases, likely driven by distinct scanning behaviors—the first phase of the task (before detection) is essentially a visual search task in which the subject fixates on visible pedestrians until an internal criterion for the potential colliding pedestrian is met leading to the subject's detection response; the second phase (after detection) then becomes a affordance task in which the subject presumably monitors the potentially colliding pedestrian while selecting a pathway that affords safe passage. Interestingly, the magnitude of the bias was larger post-detection—one possible explanation for this may be because the HH subject's residual field aligns with the vertical meridian, fixating the colliding pedestrian's center (e.g., at their eyes) would result in seeing only a portion of the pedestrian's body. As a consequence, the HH subject may bias their gaze slightly toward the blindside so that the pedestrian's entire body is within the residual visual field, thereby providing sufficient visual information so specify the collision event.

*4.5. Both normal vision and homonymous hemianopia subjects bias their vertical head rotation upwards during collision detection and avoidance.*

While we had no explicit expectations with regard to any potential biases in vertical head rotation, we did find a general bias in upward rotation in both subjects groups. One possible explanation may be that the subjects rotate their heads upward to counteract the weight of the HMD. A more likely explanation is that this bias may be driven by the relative utility of judging orientation and heading from the head and upper torso (as opposed to the legs or feet) of other pedestrians to determine the likelihood of any future collisions. This effect may be more pronounced during the avoidance maneuver where the subject is actively monitoring the pedestrian for unexpected movements that may be preceded by anticipatory head or trunk of the pedestrian. Again, we cannot infer the distribution of the subject's attention without the addition of eye tracking data. Future work will investigate the individual features of the environment and their content (e.g., orientation, social information, the pedestrian's attention, etc.) that are fixated by the subject during detection and during the avoidance maneuver.

*4.6. Conclusion*

To conclude, in this paper, we presented an HMD-based VR collision detection and avoidance testing platform that allows unrestricted walking and scanning behaviors without risk to the subject, which can successfully differentiate between PFL and NV subjects in several objective measures of mobility performance. Such a platform can easily be adopted by clinicians and other researchers interested in the assessment and rehabilitation of patients with mobility deficits by leveraging our proposed objective metrics, at low cost and effort of acquiring and operating the mobile HMD. Future work will make improvements and extend the utility of the current collision detection and avoidance task by further enhancing the ecological validity of the task demands and further developing for AR/MR platforms for use outside of the laboratory and clinical spaces.

**Acknowledgments**

This work was supported by the National Institutes of Health NEI R01-031777, P30 EY003790.

**Tables, legends, figures, supplementary material**

**Video S1. Demonstration of HMD passthrough function, realignment task, and collision avoidance trial.** The HMD's passthrough function is activated at the beginning and end of each trial. While activated, the HMD's outward-facing cameras capture and display images of the surrounding environment (shown as grayscale with edge highlighting) overlaying the rendered virtual images, both of which are shown at reduced opacity. The subject then realigns the real and virtual worlds by resetting the viewpoint of the virtual world such that the walking path (virtual green floor markers) overlaps with striped tape affixed to the real-world floor (striped central line on the floor in the passthrough view). The subject then begins a collision trial in which the colliding pedestrian approaches from $\beta = -20°$.

**Video S2. Depiction of real-world walking, perspective in the virtual environment, and an overhead view of the data stream (subject and pedestrian locations) recorded by the HMD.** The subject walks unrestricted in a large empty room while wearing the HMD and holding the remote controls. Translations and rotations of the HMD produce corresponding translations and rotations of the perspective in the virtual environment, allowing the subject to walk through the virtual mall while avoiding virtual pedestrians. The HMD records the locations and trajectories of the subject and pedestrians in real time, allowing for the visualization of the subject's avoidance behaviors in the presence of the other pedestrians.